\documentclass[useAMS,usenatbib,usegraphicx]{mn2e}

\newcommand{\kev}{keV}

\newcommand{\fe}{Fe~K$\alpha$}
\newcommand{\etal}{et al.}
\newcommand{\mcg}{MCG--6-30-15}
\newcommand{\oeight}{O~\textsc{viii}}
\newcommand{\osev}{O~\textsc{vii}}

\newcommand{\nsix}{N~\textsc{vi}}
\newcommand{\csix}{C~\textsc{vi}}

\def\deg{^{\circ}}
\def\sax{{\it BeppoSAX}}
\def\gs{\mathrel{\hbox{\rlap{\hbox{\lower4pt\hbox{$\sim$}}}\hbox{$>$}}}}
\def\ls{\mathrel{\hbox{\rlap{\hbox{\lower4pt\hbox{$\sim$}}}\hbox{$<$}}}}

\title[Ionized reflection model of MCG--6-30-15]
  {A two-component ionized reflection model of MCG--6-30-15}
\author[D.\ R.\ Ballantyne \etal]
  {D.~R.~Ballantyne,$^{1,2}$\thanks{ballantyne@cita.utoronto.ca}
  S.~Vaughan,$^2$ and A.~C.~Fabian$^2$\\
  $^1$Canadian Institute for Theoretical Astrophysics, 60 St. George
  Street, Toronto, Ontario, Canada M5S 3H8 \\
  $^2$Institute of Astronomy, Madingley Road, Cambridge CB3 0HA}

\pagerange{\pageref{firstpage}--\pageref{lastpage}}
\pubyear{2003}

\usepackage{times}

\begin{document}

\label{firstpage}

\maketitle

\begin{abstract}
Ionized reflection has often been considered as the explanation for
the unusual \fe\ variability observed in \mcg. In this paper, we test
this model using a 325~ks observation of \mcg\ by \textit{XMM-Newton}
and \textit{BeppoSAX}. The data are fit between 2.5 and 80~\kev\ with
the constant density models of Ross \& Fabian. The best fit ionized
reflection model requires the \fe\ line to be split into two
reprocessing events: one from the inner disc to build up the red wing,
and the other from the outer accretion disc to fit the blue horn. The
implied geometry is a disc which becomes strongly warped or flared at
large radii. A good fit was obtained with a solar abundance of iron
and a reflection fraction ($R$) of unity for the inner
reflector. The combination of the two reflection spectra can
appear to have $R>2$ as required by the \textit{BeppoSAX} data. The
inner reflector has an ionization parameter $\log \xi =3.8$, but the
outer one is neutral with an inner radius $\sim 70$ gravitational radii
($r_g$), corresponding to a light crossing time of about an hour for a
10$^7$~M$_{\odot}$ black hole. Applying this model to time-resolved
spectra shows that the inner reflector becomes more ionized as the
source brightens. This reduces the strength of the red wing at high
flux states. The X-ray source is constrained to arise
from a narrow annulus at $\sim 5$~$r_g$, with only 6 per cent of the
2--10~\kev\ flux due to the outer reprocessor. This amount of
localized energy generation is extremely difficult to produce without
resorting to other energy sources such as the black hole spin. In
fact, all the \fe\ models fit to \textit{XMM-Newton} spectra of \mcg\
require a large increase in energy production at the inner edge of the
accretion disc.
\end{abstract}

\begin{keywords}
galaxies: active -- galaxies: Seyfert: general -- galaxies:
individual: \mcg\ -- X-rays: galaxies
\end{keywords}

\section{Introduction}
\label{sect:intro}
X-ray observations of active galactic nuclei (AGN) allow for an
unparalleled probe of the physics at the inner edge of the accretion
disc. Spectral features such as a strong \fe\ line at 6.4~\kev\ (see
\citet{rn03} for a recent review)
and a flattening of the power-law slope above $\sim$10~\kev\ \citep{pou90,np94}
are indications that some fraction of the X-radiation is being
reprocessed (or reflected) in optically-thick material
\citep*[e.g.,][]{gf91,mpp91}. Using the \textit{ASCA} observatory
\citet{tan95} discovered a broad, asymmetric \fe\ line in the bright
Seyfert~1 galaxy \mcg\ ($z=0.008$) that was consistent with the
profile expected from an emission line produced near 6 gravitational
radii (6~$r_g$; $r_g=GM/c^2$) from the black hole
\citep{fab89,lao91}. Therefore, the reflection signatures observed in
X-ray spectra of AGN are likely caused by reprocessing within the
accretion disc -- a plausible scenario if the X-ray source resides in
a hot corona above the disc \citep*{gal79,haa91,haa93}.

In the disc-corona geometry it is expected that the \fe\ line will
respond rapidly ($\sim 50 r_g$~s for a 10$^7$~M$_{\odot}$ black hole)
to any change in the continuum \citep{stel90,mp92}. Thus, the
expectation is to observe a correlation between the line flux and the
continuum flux, and a constant \fe\ equivalent width (EW). However,
many long observations of \mcg\ have shown that the Fe line exhibits
little variability \citep*{lee00,rey00,sif02,fab02} despite large
(factors of 2--3) changes in the continuum flux, and an EW that
decreases as the source brightens. Moreover, when variations were
detected they were found to be uncorrelated with the continuum
variability \citep{iwa96,iwa99,ve01}. This perceived lack of \fe\
reverberation is a major challenge to the disc-corona model of X-ray
production in AGN.

A possible solution to this problem is if the Fe line arises from the
surface of a strongly photoionized accretion disc
\citep{iwa96,lee00,rey00,br02}. When iron is highly ionized then
increasing the strength of the illumination will result in a weaker
\fe\ line relative to the continuum (see Fig.~2 in
\citealt*{mfr96}). The flux of the line can thus remain approximately
constant despite variations in the driving continuum. This situation
can occur when helium-like Fe is the dominant ionization state
\citep*{mfr93,bfr02}, and thus requires the disc surface to have a
high ionization parameter, $\log \xi > 3$, where $\xi=4\pi
F_{\mathrm{X}}/n_{\mathrm{H}}$. The major difficulty with this
solution is that the He-like \fe\ line occurs at 6.7~\kev, but the
line core of \mcg\ is well established to be at 6.4~\kev, and does not
arise from very distant material \citep{lee02}. Therefore, at first
glance it seems unlikely that the \fe\ line arises from an ionized
disc.

One way to check this interpretation is to compare the entire spectrum
to models of X-ray reflection from accretion discs. Calculations of X-ray
reprocessing have shown that the shape of the spectrum is sensitive to
the ionization state of the material \citep*{ros93,ros99}, its density
distribution \citep*{nkk00,brf01,dum02,roz02}, and the properties of the
illuminating radiation \citep{br02}. These models self-consistently
calculate the strength of the \fe\ line which changes as a function of
ionization state and abundance \citep{bfr02}, so the line and the entire
continuum can be fit simultaneously with only a few parameters. 

This exercise was previously performed for \mcg\ by \citet{bf01} using
\textit{ASCA} data. They found that the data between 3 and 10~\kev\
(the 3~\kev\ cut-off was used to avoid the warm absorber) was well fit
by an ionized reflector, but required the addition of a separate line
component further out along the disc. However, a good fit was also
obtained by a single, neutral constant density reflector with twice
the solar iron abundance. From those data it was unclear which
interpretation was more plausible.

Producing a physical model for the X-ray spectrum and spectral
variability of \mcg\ is highly involved and there is no unique
solution to this problem. In this paper we explore the implications
for models in which the spectrum is dominated by features from a
photoionized accretion disc.  Specifically, we improve on the results
of \citet{bf01} and develop a model for the spectrum of \mcg\ in terms
of reflection from two distinct sites on the disc and test this model
against the observed spectral variability.  An alternative model, in
which the reflection spectrum is physically disconnected from the
continuum due to relativistic effects, is discussed by
\citet{fv02}. We make use of data obtained from a 325~ks observation
of \mcg\ by \textit{XMM-Newton} and from a simultaneous observation by
\textit{BeppoSAX} \citep{fab02}.

In the next section, the data and models are briefly introduced. In
Section~\ref{sect:sax} we use the broad-band \textit{BeppoSAX} data to
constrain the reflection spectrum in a manner independent of the iron
line. Section~\ref{sect:res} presents the results of the ionized disc
fits. The findings are discussed in Section~\ref{sect:discuss} before
conclusions are presented in Section~\ref{sect:concl}.

\section{Data and Models}
\label{sect:data}
Details of the observations are given in \citet{fab02}.  For the
present paper the data were processed using \textit{XMM-Newton}
Science Analysis System (SAS) v5.3.3, with source data extracted from
circular regions of radius 35$^{\prime\prime}$ from the EPIC MOS and
pn. Background events were
extracted from off-source regions.  Events corresponding to patterns
0--12 (single--quadruple pixel events) were extracted from MOS and
patterns 0--4 (singles and doubles) were used for the pn analysis,
after checking for consistency with the data extracted using only
single pixel events (pattern 0). Standard redistribution matrices
(\texttt{m1\_r6\_all\_15.rmf}) for MOS1 (similarly for MOS2) and
(\texttt{epn\_sw20\_sdY9.rmf}) for the pn were used, and ancillary
response files were generated using \textsc{arfgen v1.48.8}.

The additional calibration information included in this version
of the SAS now allow the use of the pn data (in \citet{fab02}
only the MOS data was analyzed). However, it was
found there was a small but significant difference in photon-index
($\Delta \Gamma \sim 0.1$) between the MOS and pn spectra even after
removing the small amount of pile-up in the MOS data. Since
the value of $\Gamma$ obtained from the pn data was more
consistent with that from the MECS and PDS instruments, we ignored
the MOS data in this analysis (see \citet{fv02} for further discussion
on the EPIC data).

The constant density ionized reflection models of \citet{ros93} and
\citet{ros99} were used in the spectral fitting, as these are the most
useful in parameterizing the shape of the spectrum. Models where the
accretion disc atmosphere is in hydrostatic balance
\citep[e.g.,][]{nkk00,brf01,roz02} were not used because they are dependent
on the unknown accretion disc structure. However, \citet{brf01} found
that the hydrostatic models could be fit with dilute versions of the
constant density ones. Therefore, there is not a significant difference in
the shape of the reflection spectrum between the two cases.

\textsc{xspec} v11.2 \citep{arn96} was used for all spectral fitting,
and we use the 90 per cent confidence level to compute errorbars for
the fit parameters.

\section{The Reflection Strength}
\label{sect:sax}

As a first step toward determining the properties of the reflection
spectrum in \mcg, the high-energy \sax\ PDS data were examined in
order to constrain the strength of the Compton reflection ``hump.''
In this analysis models of the reflected continuum were compared to
the PDS data over the 14--200~keV range, which includes the peak of the
reflection spectrum.  The MECS data were included in the fitting to
better constrain the slope of the underlying continuum. Only MECS data
in the ranges 2.5--3.5~keV and 7.5--10.0~keV were used as these should
be dominated by the primary X-ray continuum;  the spectrum is
complicated by absorption below 2.5~keV and  the strong, broad iron
line in the $\sim 3$--$7$~keV range. The normalization of the PDS data
was tied to 0.86 times the MECS value \citep*{fgg99}.
The EPIC data were not included
in these fits as the much higher signal-to-noise ratio of these data
would mean the model parameters would be determined almost entirely by
the low-energy EPIC data, not the PDS.

\begin{figure}
\centerline{
\includegraphics[angle=-90,width=0.5\textwidth]{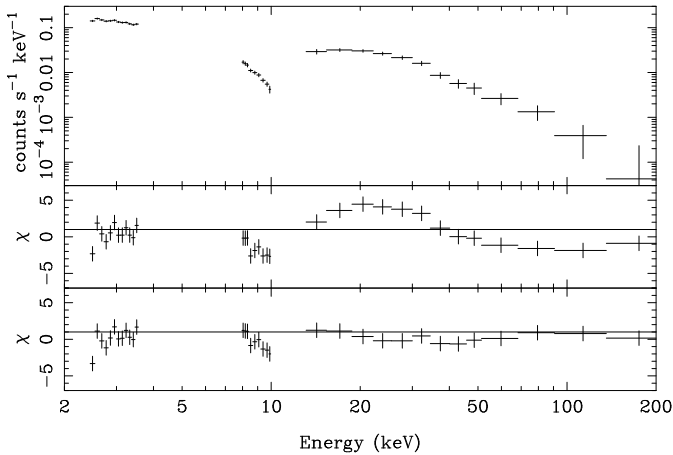}
}
\caption{Top panel: \sax\ PDS and MECS data. Middle panel: residuals from a fit
using an exponentially cut-off power-law model. Bottom panel: residuals
allowing for reflection as described in Section~\ref{sect:sax}.}
\label{fig:saxfit}
\end{figure}

A simple power-law model for the broad-band continuum provided a poor
fit to the data ($\chi^{2}=170.7/46$ degrees of freedom, d.o.f.), with
a photon index of $\Gamma = 1.85$. The residuals from this fit showed
the PDS data were systematically in excess of this model in the
14--40~keV range, a strong indication of the presence of an additional
Compton reflection continuum, which peaks at $\sim 30$~keV. Allowing
for an exponential cut-off in the power-law continuum did not improve
the fit ($\chi^2=170.7/45$ d.o.f.) and gave a best-fitting cut-off
energy $E_{\rm cut} > 500$~keV, i.e. above the band-pass of the data,
implying that the high energy spectral curvature is not the result of
a roll-over in the power-law continuum. This fit is illustrated in
Fig.~\ref{fig:saxfit}. The earlier \sax\ observation reported by
\citet{gua99} also clearly detected the reflection continuum of \mcg.

The broad-band spectrum was therefore modeled in terms a power-law
plus Compton reflection continuum.  The \textsc{pexrav} model
\citep{mag95} was used to compute the spectrum. Specifically, this
computes the spectrum resulting from an exponentially cut-off
power-law continuum incident on a slab of optically thick, neutral
material, accounting for Compton reflection and bound-free
absorption. Unlike the \citet{ros93} ionized disc model,
\textsc{pexrav} does not account for discrete emission features
(specifically, fluorescence and recombination emission). However, it
does account for the angular dependence of the reflected spectrum and
the reflection of incident photons with energies
$\gs100$~keV (which requires relativistic corrections to the
scattering cross-section; \citealt*{wlz88}). This model should
therefore be appropriate for the energy ranges included in the
fitting, which extend above $100$~keV and ignore regions containing
strong atomic emission features.

Including a \textsc{pexrav} reflection component in the model improved
the fit substantially ($\chi^2=71.7/44$ d.o.f.) and gave the following
fit parameters: $\Gamma=2.17\pm0.10$, $E_{\rm cut}>143$~keV and a
normalization of the reflected spectrum relative to the primary
continuum of $R=2.7_{-0.9}^{+1.4}$, where $R=\Omega/2\pi$. The
inclination angle of the reflector was fixed at $i = 30 \deg$
\citep{tan95,fab02} and the abundances were assumed to be solar. This
fit suggests rather strong reflection; an isotropic continuum source
illuminating a flat accretion disc would be expected to yield a
relative reflection strength of $R \sim 1$ (i.e. the reflector
subtends a solid angle $\Omega \sim 2\pi$ as seen by the continuum
source).

The exact strength and shape of the observed reflection spectrum does
however also depend on the abundances of the reflector, its
inclination and and its ionization state. The effects of these
parameters on the measured value of $R$ were investigated as follows.
The \textsc{pexrav} model was re-fitted to the data with the Fe
abundance fixed at $3\times$ and then $7\times$ solar and also with
the Fe abundance as a free parameter. The $3\times$ solar model made
the largest improvement to the fit (compared to the solar case) and
the best-fitting abundance value was $3.5_{-1.9}^{+1.4}$. This gave
$\chi^2=63.6/44$ d.o.f. and the reflection fraction was constrained to
be $R>2.6$ (this fit is shown in the bottom panel of
Fig.~\ref{fig:saxfit}). Allowing the abundances of elements lighter
the Fe to vary did not improve the fit further. The Fe abundance was
therefore fixed to be $3\times$ solar in the remaining \textsc{pexrav}
models.  \citet{lee99} previously presented evidence for an enhanced
iron abundance in \mcg.

The apparent strength of the reflected spectrum, for a given value of
$R$, is also dependent upon the inclination angle of the reflector
with respect to the line-of-sight \citep[e.g., Fig.~5
of][]{mag95}. Allowing the inclination angle to be a free parameter
gave only a small improvement in the quality of the fit ($\chi^2 =
63.4 / 43$ d.o.f.) with $i < 80 \deg$ and similar constraints on the reflection
strength $R>2.8$.

The final possibility explored to explain the high value of $R$ was
ionization of the reflector.  This was performed by using the
\textsc{pexriv} model and allowing the ionization parameter $\xi$ to
vary (with $3\times$ solar Fe and $i=30 \deg$). This gave only a
slight improvement in the fit ($\chi^2 = 61.8 / 43$ d.o.f.) with $\xi
< 200$ and $R>2.8$ (this $\xi$ has been corrected for the different
energy ranges used in \textsc{pexriv} and the \citet{ros93}
model). The low value of $\xi$ means that the lack of Comptonization
of the Fe K edge in this model will not affect the limit on
$R$. Thus, even allowing for the possible effects of non-solar
abundances, angular-dependence and ionization on the reflection
spectrum, the strength of the reflection measured using the
high-energy PDS data appeared to be $R\gs 2$. This constraint on $R$
is independent of the possible presence of a roll-over in the
power-law continuum, since the cut-off energy of the power-law was
a free parameter in the above fits.

\section{Ionized Reflection Results}
\label{sect:res}
The \sax\ constraint on $R$ is difficult to explain in terms of the
standard disc-corona geometry which predicts $R \sim 1$. A large
reflection strength can be generated if the X-ray emitting plasma is
moving toward the accretion disc at mildly relativistic velocities
\citep{rf97,bel99,mbp01}. However, this model would predict an extremely soft
$\Gamma > 2.2$ spectrum for \mcg, which is not observed. Another
method of obtaining a large value of $R$ was given by \citet{fab02a}
who considered the results of reflection from a lumpy or
corrugated accretion disc (see also \citealt{rfb02}). In this
scenario, the optically thick disc may subtend a solid angle larger
than 2$\pi$ as seen from the X-ray source. Again, softer spectra are
naturally predicted from such geometry, as the hard X-ray source `sees'
more UV photons from the disc. 

When fitting the \textit{ASCA} spectra of \mcg, \citet{bf01} found
that it was very difficult for an ionized \fe\ line to fully account
for both the strength and width of the observed feature. On the other
hand, an ionized reflection continuum provided a good fit to the
observed one, including the red wing of the \fe\ line. Thus,
\citet{bf01} were forced to consider a two component model, where the
red wing was fit by the ionized reflector, and the line core by a
neutral diskline at 6.4~\kev. \citet{mis01} has also considered a two
component model of the \mcg\ iron line. A superposition of more than
one reflection event may also account for the observed reflection
strength. In this section, we investigate the validity of such a model
with our high signal-to-noise \textit{XMM-Newton} data.

\subsection{The time-averaged spectrum}
\label{sub:avg}
\mcg\ has a strong warm absorber that significantly affects the
continuum below 2--3~\kev\ \citep{np92,rey95,lee01,sak02}. Therefore, the
use of the full pn spectrum would necessitate a complete model of
the warm absorber from analysis of the Reflection Grating Spectrometer
(RGS) data. However, the strongest reflection features are not
affected by the warm absorber, so, following \citet{bf01} and
\citet{fab02}, we only considered data above 2.5~\kev. The MECS and
pn data were cut-off at 10 and 12~\kev, respectively. The PDS data
were cut-off at 80~\kev\ because the \citet{ros93} models have a high
energy limit of 100~\kev. All fits included the Galactic absorbing
column of $4.06 \times 10^{20}$~cm$^{-2}$ \citep*{ewl89}. Data from the
three instruments were fit simultaneously with the normalizations free
except for the PDS which was fixed to be 0.86 times the MECS value as
in Sect.~\ref{sect:sax}.

The model consisted of the sum of two reflection spectra with solar
iron abundance. The first of these components was relativistically
blurred with a \citet{lao91} kernel that had a power-law emissivity
($\propto r^{-\beta}$) with the index frozen at $\beta=3$. The second
component was also blurred, but with a flatter index of $\beta=2$
appropriate for the case when the disc is warped and emission from
large radii is non-negligible. Both reflectors were illuminated by the same
power-law spectrum, but this primary emission was included only in the
first component where it was added in so that $R$ was unity. Thus, the
second component was reflection dominated. In this way, a total
observed $R > 2$ can be obtained, as required by the \sax\ data
(Sect.~\ref{sect:sax}). Following \citet{fab02} a Gaussian with zero
width was included at 6.9~\kev\ to fit an additional spectral feature.

\begin{figure}
\centerline{
\includegraphics[angle=-90,width=0.5\textwidth]{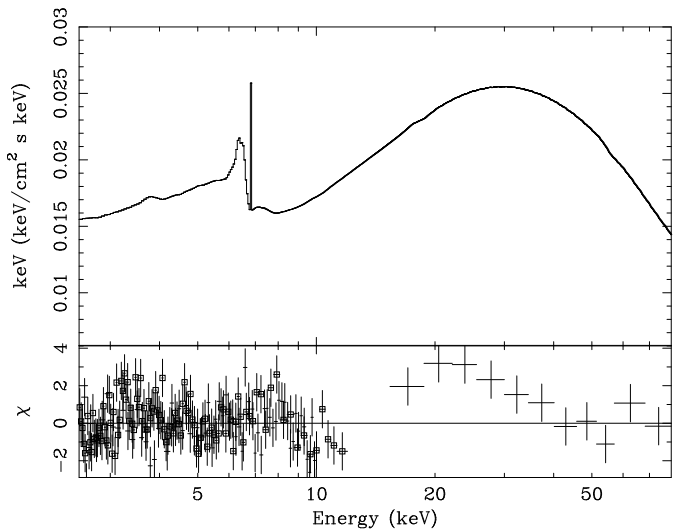}
}
\caption{The summed double reflector model fit to \mcg\ between 2.5
and 80~\kev. The top panel shows the total spectrum in $EF_E$ space
(see Figure~\ref{fig:dblfs} for a plot showing the model components),
and the lower panel are the residuals in units of standard
deviations. The pn data are shown with the open symbols, and the fit
parameters are displayed in Table~\ref{table:dbldata}. The residuals
at 20--30~\kev\ are a result of the reflection spectra being
angle-averaged, which reduces the strength of the reflection hump
\citep{mag95} for inclination angles $< 50$~degrees.}
\label{fig:dblfit}
\end{figure}

The best fit obtained with this model is shown in
Figure~\ref{fig:dblfit} and Table~\ref{table:dbldata}.  
We find a very good fit to the data ($\chi^2/$d.o.f.=1600/1879) with an
ionized reflector located at about 5~$r_g$ and a neutral reflector
outside of $\sim$70~$r_g$. The photon index is consistent with the
results of \citet{fab02} ($\Gamma=1.95$; see their Table~1). The inner
radius found here ($\sim 5$~$r_g$) is farther out than that found
by \citet{wil01} or \citet{fab02} ($\sim 2$~$r_g$). This is because
the ionized reflection model has both a curved spectrum and an
intrinsically broadened \fe\ line (due to Comptonization). The
ionization parameter of the inner reflector ($\log \xi = 3.82$) is
such that He-like iron is the dominant ion at the surface of the
disc. Therefore, the rest energy of the \fe\ line is at 6.7~\kev, but is
gravitationally redshifted down to below 6~\kev.

Examining the residuals to this fit in Fig.~\ref{fig:dblfit} shows
that the reflection hump is not accurately reproduced. This problem is
due to the fact that the \citet{ros99} reflection models are
angle-averaged rather than computed for a specific inclination
angle (as is done in \textsc{pexrav}). For angles $< 50$~degrees,
angle-averaging results in a weaker reflection hump at 20--30~\kev\
\citep{mag95}. A possible secondary effect
is the hard 100~\kev\ cut-off in the \citet{ros99} models causing the
spectrum to turn downwards faster than is observed. However, cutting
off the PDS data at lower energies (say, 40--50~\kev) makes little
impact on the model parameters. Despite these issues,
Fig.~\ref{fig:dblfit} illustrates that more than one reprocessing
event can account for the strong reflection hump observed in the \sax\ data. 

Allowing $\beta$ for the inner reflector to be a free parameter does
not improve the fit for any value between 10 and $-$10 because the
inner and outer radii are so close together (see below). Doing the same
for the outer component improves the fit only slightly with $0.641
\leq \beta_{\mathrm{out}} \leq 2.268$. This result emphasizes that
the model requires a flat emissivity at large radii, consistent with
a strongly warped or concave disc. Allowing the reflection
fraction to be free does drop the $\chi^2$ by 8, which is significant
at 99.8\% according to the F-test. The best fit value of $R$ in this
case is 1.98$^{+0.97}_{-0.44}$. This additional reflection only
affected the fit around the \fe\ line, and did not alter the residuals
at higher energies.

\begin{table*}
\begin{minipage}{170mm}
\caption{The parameters for the best fit time-averaged double
  reflection model shown in Figure~\ref{fig:dblfit}. $\Gamma$ is the
photon-index of the incident power-law continuum, $\xi$ is the
ionization parameter of the reflector, $r_{\mathrm{in}}$ is the inner
radius of the reflector in $r_g$, $r_{\mathrm{out}}$ is the outer
radius (also in $r_g$), and $i$ is the inclination angle in
degrees. The EWs of the 6.4~\kev\ line and the narrow 6.9~\kev\ line
are given in eV. The EW of the broad \fe\ line includes both
components to the line profile. The inner reflector had $R$ fixed at
unity, and an emissivity of $\beta=3$. The outer component was
reflection dominated and had an emissivity of $\beta=2$. $_p$ means
the parameter pegged at its lower limit.}
\label{table:dbldata}
\begin{tabular}{ccccccccccc}
 & \multicolumn{3}{c}{Inner reflector} & \multicolumn{3}{c}{Outer
reflector} & & & & \\
$\Gamma$ & $\log \xi$ & $r_{\mathrm{in}}$ & $r_{\mathrm{out}}$ & $\log
\xi$ & $r_{\mathrm{in}}$ & $r_{\mathrm{out}}$ & $i$ &
EW$_{\mathrm{6.4\,keV}}$ & EW$_{\mathrm{6.9\,keV}}$ &
$\chi^2/$d.o.f.\\ \hline 
1.92$^{+0.02}_{-0.01}$ & 3.82$^{+0.02}_{-0.03}$ & 4.93$\pm 0.01$ &
5.012$^{+0.003}_{-0.002p}$ & $<1.07$ & 71$^{+64}_{-33}$ &
1822$^{+1425}_{-732}$ & 31.6$^{+0.8}_{-1.0}$ & 450 & 29$^{+31}_{-5}$ &
1600/1879 \\
\end{tabular}
\end{minipage}
\end{table*}
In this model, the large red wing of the Fe line and the 2--10~\kev\
continuum are due to the inner ionized reflector, while the blue core
of the line and the reflection hump is dominated by the outer
reprocessor. The most striking feature of this fit is that the radii over
which the inner component originates is restricted to a narrow annulus
only 0.08~$r_g$ wide. In Sect.~\ref{sub:impli} we discuss the
implications of this model and attempt to evaluate if it is physically
plausible. 

\subsection{Time resolved spectra}
\label{sub:variability}

\begin{figure*}
\begin{minipage}{180mm}
\centerline{
\includegraphics[angle=-90,width=\textwidth]{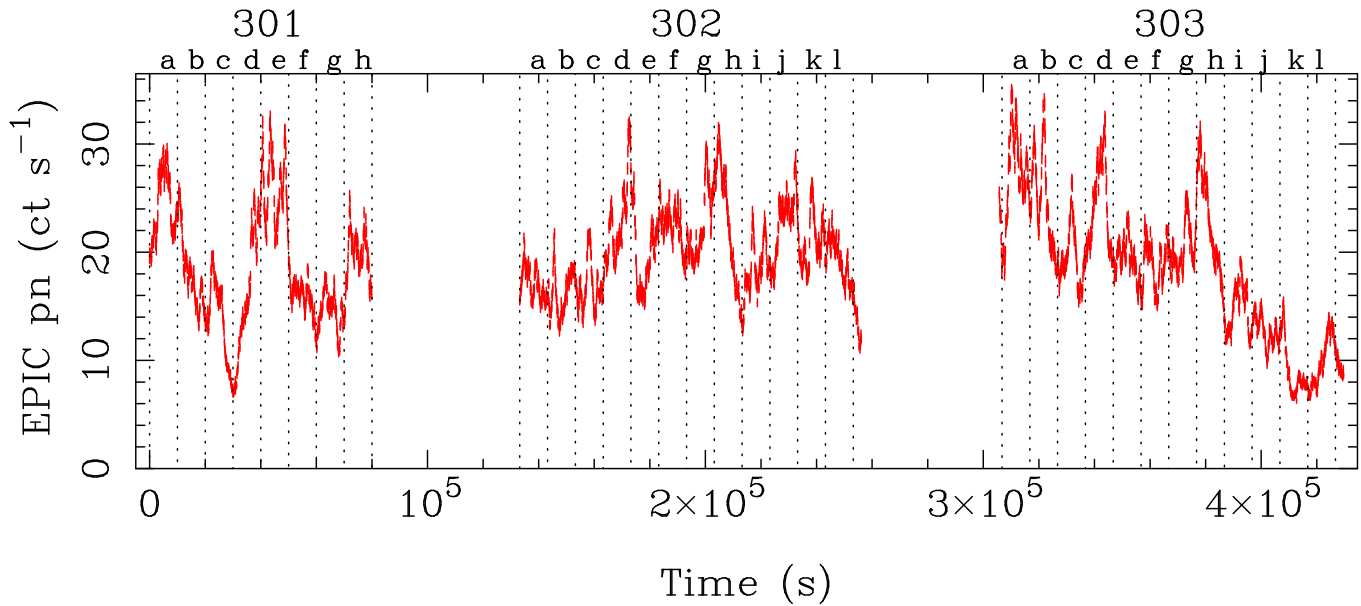}
}
\caption{\textit{XMM-Newton} EPIC pn light curve (0.2--10~\kev) of
  \mcg. The dotted lines denote the 10~ks segments discussed in the
  text. Source and background spectra were extracted from each segment.
  The total good exposure time of each spectrum was
  $\sim$7.1~ks due to the live-time of the pn in small window
  mode. The count rates in this figure have not been corrected for
  this scale factor.} 
\label{fig:lc}
\end{minipage}
\end{figure*}

In order to investigate whether the double reflector model can explain
the spectral variability properties of \mcg, we split the
\textit{XMM-Newton} dataset into 32 10~ks segments (see Figure~\ref{fig:lc}).
Source and background spectra were extracted from the pn for each
interval. Due to the $\sim$71\% observing efficiency of the pn
\citep{stru01}, the exposure time of each spectrum was $\sim$7.1~ks.  

Each of the 32 spectra were fit with the double reflection model
between 2.5 and 12~\kev\ using the time-averaged results
(Table~\ref{table:dbldata}) as a template. All parameters were allowed
to float, except for the inner and outer radii of the distant
reflector, the reflection fraction $R$, and the inclination angle $i$
which were all fixed at the values listed in
Table~\ref{table:dbldata}. Keeping $R$ fixed allows us to test whether
the data are consistent with the innermost reflection component
varying (in flux) along with the continuum (allowing for changes in
the ionization parameter). The emissivity indices also remained
unchanged from the time-averaged fitting. The model provided an
acceptable fit (reduced $\chi^2 \sim 1$) to the continuum and \fe\
line for each spectrum.

\begin{figure*}
\begin{minipage}{180mm}
\centerline{
\includegraphics[width=0.5\textwidth]{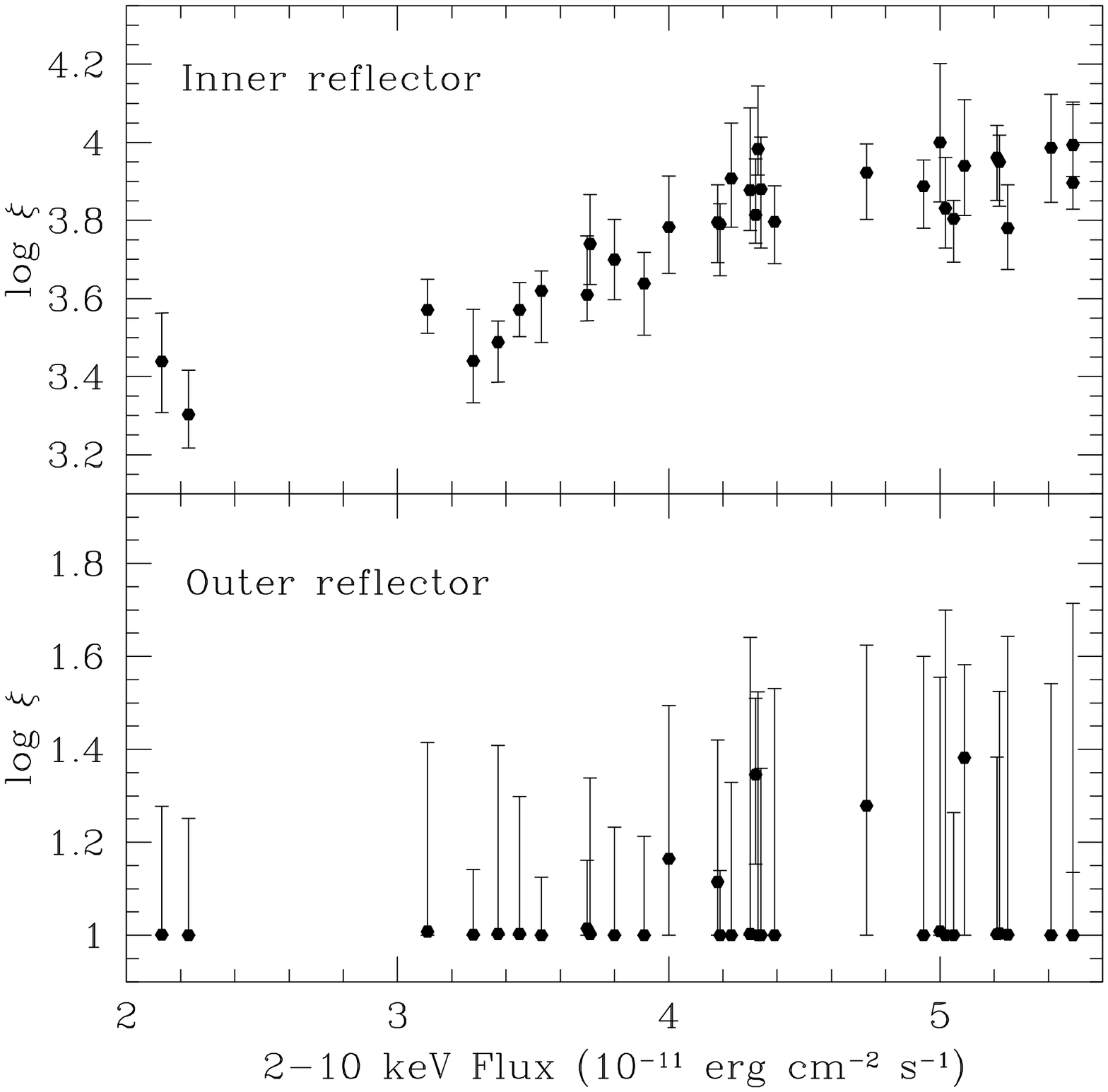}
\includegraphics[width=0.5\textwidth]{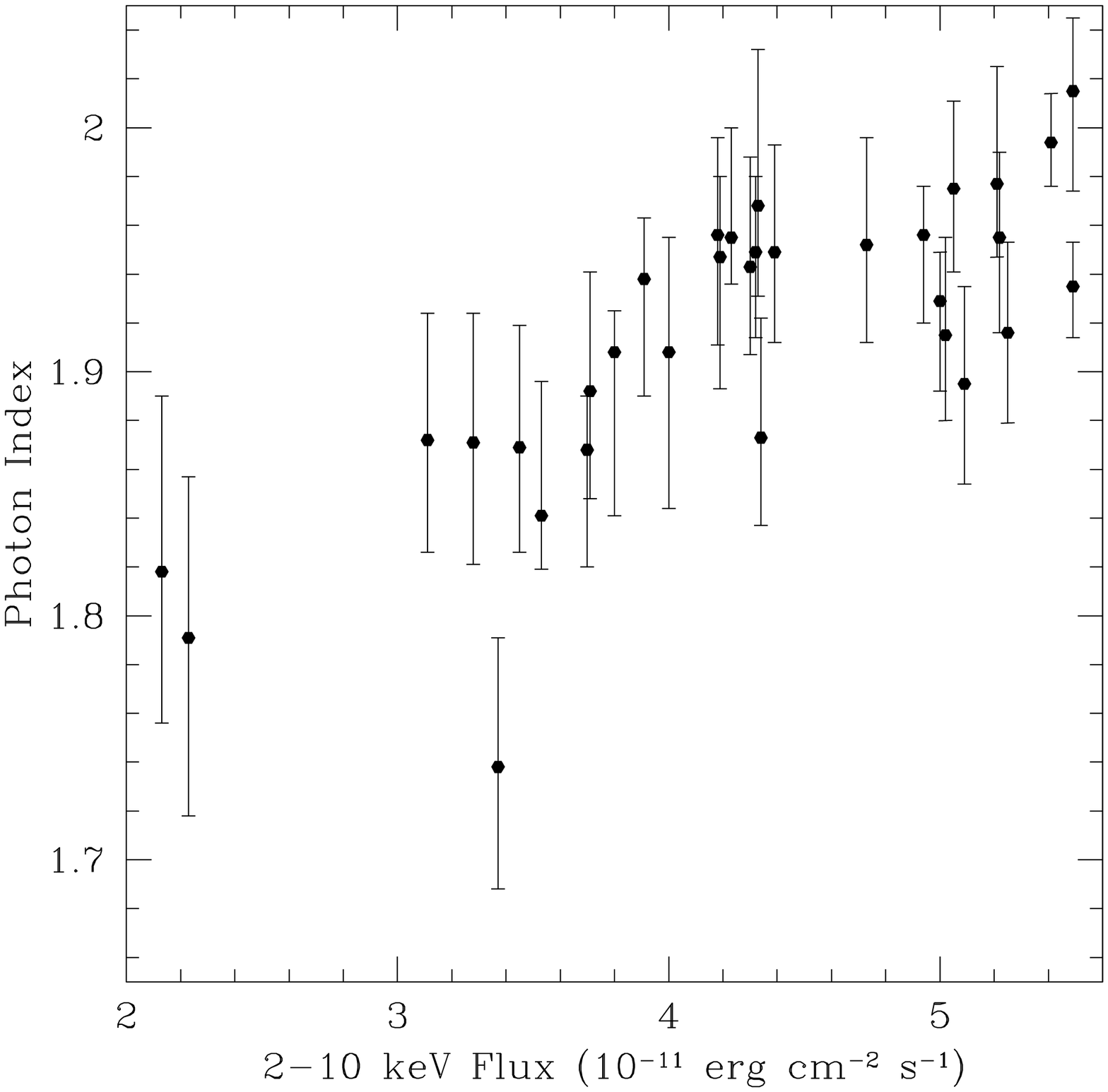}
}
\caption{\textit{Left}: Plot of the ionization parameter for the two
reflectors versus the total 2--10~\kev\ flux. The inner reprocessor clearly become
more ionized as the source becomes brighter, while the distant
reflector remains neutral. \textit{Right}: Plot of the photon index
($\Gamma$) versus the total 2--10~\kev\ flux. The source becomes softer as it
brightens, which is in agreement with many other observations of \mcg\
using simpler models.} 
\label{fig:xi-gam-cr}
\end{minipage}
\end{figure*}
The results from this fitting are shown in
Figure~\ref{fig:xi-gam-cr}. The left panel shows the ionization
parameter of the two reflectors as a function of the total 2--10~\kev\
flux calculated from the model. This plot clearly shows that the inner
reprocessor at $\sim$5~$r_g$ becomes more ionized as \mcg\ becomes
brighter. The relationship between $\xi$ and the flux is roughly
linear over most of the plot. The red wing of the \fe\ line (which is
fit by this reflector) will therefore be weak at high flux levels. In
contrast, the ionization parameter of the outer reflector is
uncorrelated with the 2--10 \kev\ flux and remains small, consistent
with predominantly neutral reflection. The right panel in
Fig.~\ref{fig:xi-gam-cr} plots $\Gamma$ as a function of 2--10~\kev\
flux. Here, we see that we have recovered the well known property that
the spectrum of \mcg\ softens as it becomes brighter
\citep[e.g.,][]{ve01,sif02}.

\section{Discussion}
\label{sect:discuss}
\textit{XMM-Newton} has observed \mcg\ twice. The first observation
showed the broad iron line likely extends down below 4~\kev, implying
emission from within $6$~$r_g$, and also showed a narrower core to the
line \citep{wil01}. This apparent two component line may suggest that
there is reflection from two distinct sites. The second reprocessing
site is unlikely to be very distant (e.g., the molecular torus
postulated in Seyfert unification schemes) as the high resolution
\textit{Chandra} HETGS spectrum showed little flux from an
intrinsically narrow line component \citep{lee02}. The second, longer
\textit{XMM-Newton} observation also appeared to show line emission
extending down to low energies. \citet{fab02} modeled this in terms of
neutral reflection from an accretion disc with a broken power-law
emissivity profile. In this model the extended red wing of the line is
produced by emission within $6$~$r_g$, while the core of the line
profile (which was resolved) is produced from reflection beyond
$6$~$r_g$ and with a relatively flat emissivity profile. These
analyses both suggest that the core of the line around 6.4 keV is
mostly produced by reflection off material at distances of $\sim 6 -
1000$~$r_g$, while the red wing of the line is produced by emission
from within $6$~$r_g$. Such an interpretation is consistent with the
ionized reflection results found above. In this section we consider
the two component ionized disc model in detail.

\subsection{General comments}
\label{sub:gen}
In Section~\ref{sect:res} it was found that ionized reflection models
can fit the \textit{XMM-Newton} data of \mcg\ but only with some
difficulty. As was found previously \citep{bf01}, it is difficult for
the predicted spectrum to account for the strength and width of the
\fe\ line, as well as the smooth continuum. The most tenable model
required a secondary reflector from farther out on the disc, but, more
worrying, the region of primary X-ray generation is confined to a very narrow
annulus. Are there limitations in our reflection spectra which is
resulting in these extreme models? Although it is difficult to answer
this question definitively because the detailed physics of AGN is
relatively unknown, the \citet{ros93} reflection models have been very
successful in fitting the X-ray spectra of many different Seyferts
\citep*{bal01,orr01,der02,dfp02}. On the computational side, the code
has been compared with similar angle-averaged reflection calculations
\citep{peq01} and there was good agreement at hard X-rays,
particularly around the \fe\ line. The models used here assume
reflection from a constant density slab, rather than a variable
density atmosphere, as was done in \citet{bf01}. In these variable
density models weaker \fe\ lines are expected due to the smaller
density at the surface of the disc; therefore, a constant density
slab, although perhaps unrealistic, is the ``best case scenario'' for
producing strong \fe\ lines. Thus, it seems unlikely that this extreme
fit is a result of limitations in the reflection models.

It is worth pointing out that, consistent with the results of
\citet{wil01} and \citet{fab02}, the \fe\ line profile in this model
suggests a concentration of emission from within $\sim$5~$r_g$. It is
unknown whether this is from the plunge region of a Schwarzschild
black hole, or from the disc around a spinning Kerr hole. In either
case, it seems as though additional physics must be included into our
thinking of energy generation in AGN.

\subsection{Implications of a double reflection model}
\label{sub:impli}

\subsubsection{Observational testable predictions}
\label{subsub:obs}
\begin{figure}
\centerline{
\includegraphics[angle=-90,width=0.5\textwidth]{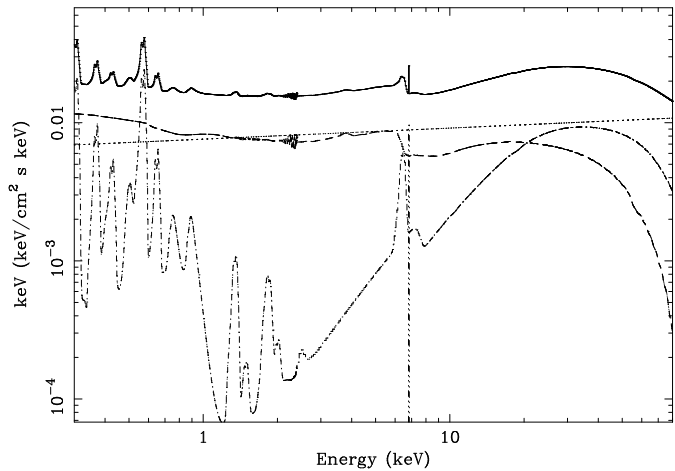}
}
\caption{The time-averaged reflection spectrum between 0.3 and 80~\kev\
predicted by the double reprocessor model discussed in
Sect.~\ref{sub:avg}. Galactic absorption has been removed in order
to show the soft X-ray lines predicted by the model. The solid line is
the full model; the dashed line denotes the inner, ionized reflector;
the incident $\Gamma=1.92$ power-law is shown with the dotted line,
and the outer, neutral reprocessor is the dot-dashed line. The
6.9~\kev\ is also included in the plot. The noise at $\sim$2~\kev\
is a plotting artifact that did not affect the spectral fit.}
\label{fig:dblfs}
\end{figure}

In Figure~\ref{fig:dblfs} the double reflection model is broken into
its components and plotted between 0.3 and 80~\kev. Absorption from
the Galaxy has been removed so that the soft X-ray spectral features
can be seen unhindered. 
The total 2--10~\kev\ flux given by the model is $4.36\times
10^{-11}$~erg~cm$^{-2}$~s$^{-1}$, of which only 6\% is contributed by the
distant component. When the energy range is
extended to 100~\kev, the outer reflector provides 21\% of the total
flux. Thus, if this model is viable we would expect a decrease
in rapid continuum variability at higher energies, as the more distant
emission would contribute to a larger fraction of the total
spectrum. Furthermore, there may be a detectable lag in variations at
high energies compared to those occurring between 2 and 10~\kev. These
predictions could be tested by \textit{Integral} or \textit{ASTRO-E2}.

Another observationally testable aspect of this model are the soft X-ray
spectral features. The ionization parameter of the outer reflector is
constrained to be small enough for iron to be only weakly ionized, but
in these cases recombination emission from ions of highly ionized
metals such as C, N and O are common at lower energies
\citep{ros93}. As there is only an upper limit on the $\xi$ of the
outer component, we cannot predict the exact lines which may be
present in the total spectrum. However, as an example, the EWs of the
soft X-ray lines relative to the total predicted continuum are shown
in Table~\ref{table:dblews} for the case when $\log \xi = 1.0$.
\begin{table}
\begin{center}
\caption{Equivalent widths (EWs) of the soft X-ray emission lines
predicted by the double reflection model when $\log \xi=1.0$ for the
outer reflector.}
\label{table:dblews}
\begin{tabular}{cc}
Line & EW (eV) \\ \hline
Si K$\alpha$ & 6 \\
Mg K$\alpha$ & 4 \\
\oeight\ Ly$\alpha$ & 9 \\
\osev\ & 36 \\
\nsix\ & 6 \\
\csix\ Ly$\alpha$ & 9 \\
\end{tabular}
\end{center}
\end{table}
The strongest line is predicted to come from \osev\ with an EW of
$\sim$36~eV. Once an accurate model for the warm absorber has been
determined it should be possible to search for such a line in the
current dataset. The line is not extremely relativistically blurred,
so may even stand out in the RGS spectra and would be unrelated to the
relativistic soft X-ray line model of \citet{bran01}.

In Sect.~\ref{sub:variability} it was found that the ionization
parameter of the inner reflector was correlated with the 2--10~\kev\
flux and was always greater than 1000. This implies that the strength
of the iron line (which comprises the red wing of the observed \fe\
line) is anticorrelated with the flux \citep{mfr96,bfr02}. However, if
the flux of \mcg\ drops to about one-half of the faintest level
observed here so that $\log \xi < 3$, then the strength of the red
wing correlates with $\xi$. This is another observational testable
feature.

\subsubsection{Geometry}
\label{subsub:geom}
The geometry implied by the double reflector fit requires two
different changes in curvature in the accretion disc. First, the disc
must become strongly concave at a radius of $\sim$50--70~$r_g$ to
provide the surface for the second reflection component. A standard
Shakura-Sunyaev gas-pressure dominated accretion disc is already
mildly concave with the scale height $H \propto r^{9/8}$ \citep{ss73},
but a warp much larger than this is required to substantially increase
the reprocessed flux \citep[see Fig. 2 in][]{bla99}. A possible
mechanism may be warping due to radiation pressure on the disc
\citep{pri96,pri97}, however the typical transition radius for this
instability is $\sim 1000$~$r_g$ for an AGN disc \citep{pri97}. The
Bardeen-Petterson effect \citep{bp75,kp85}, which aligns the inner part of an
accretion disc so that it is perpendicular to the spin axis of the
central object, is another possible method to warp a
disc if the outer regions are not in the equatorial plane. The transition
radius for this mechanism is $\la 100$~$r_g$ \citep{ii97,np00}, so the
Bardeen-Petterson effect can account for the outer disc geometry
implied by the double reflection model. Of course, a spinning black
hole would be needed to cause this effect.  

\begin{figure}
\centerline{
\includegraphics[width=0.50\textwidth]{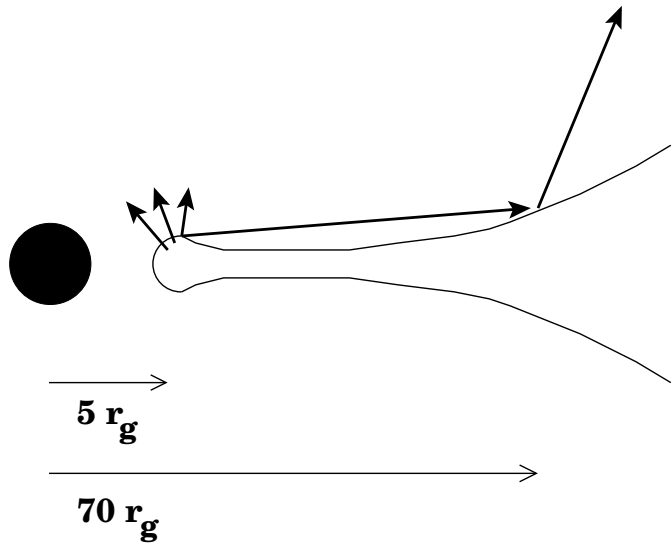}
}
\caption{A schematic representation of a possible geometry implied by
the double reflection model. X-rays are produced at the inner edge of
the accretion flow which is thicker than the rest of the inner disc
(possibly because of large radiation pressure support). Reflected
emission from this environment is greatly blurred and highly
ionized. The increased height shields the inner disc from much of the
direct radiation, and it is not until the disc becomes more concave at
large radii that it produces more reflection.}
\label{fig:schematic}
\end{figure}

Another constraint on the geometry is provided by the inner
reflector. Here, the emission is confined to a narrow ring at
$\sim$5~$r_g$ (see Sect.~\ref{subsub:annul}) and there is no
reprocessing between that point and the outer disc, ten times further
away. This implies that the inner X-ray source is hidden from most of
the inner accretion disc, which requires the scale height to rise
quickly at $\sim$5~$r_g$. A schematic diagram illustrating this
geometry is shown in Figure~\ref{fig:schematic}.
The increase in disc thickness could be due to an increase in the
radiation pressure support in the innermost region of the accretion
flow. 

\subsubsection{\fe\ variability}
\label{subsub:vary}
The most intriguing aspect of the double reflector model is its
potential to possibly explain the puzzling variability of the \fe\
line profile \citep{iwa96,iwa99}. Analysis of the long (4.5 day) 1994
\textit{ASCA} observation of \mcg\ by \citet{iwa96} considered the
variability of the line core separately from the broad red wing. They
found that on timescales greater than a few $10^4$~s, the narrow
component seemed to increase with the continuum flux, while the
strength of the broad component decreased. On shorter timescales of a
few $10^3$ seconds the broad component varied, but the narrow core was
consistent with staying constant. Reinforcing these trends was the
spectral evidence provided by examining the data during flares or dips
in the light curve. \citet{iwa96} found that during a bright flare,
the spectrum (integrated over 36~ks) showed a strong line core, but a
very weak red wing. In contrast, a 15~ks deep minimum at the end of
the observation showed a strong red wing and a weak blue core. Another
interesting line profile was found during a bright, hour long flare in
the 1997 long \textit{ASCA} observation of \mcg. \citet{iwa99}
reported that the spectrum extracted from the flare had line emission
only from within 5~$r_g$. There was little evidence for a core at
6.4~\kev, with a limit on the EW of 60~eV.

A long 400~ks \textit{RXTE} observation also illustrated the strange
behaviour of the \fe\ line in \mcg. Although the PCA on \textit{RXTE}
could not resolve the line into a core and wing, these data were used
by \citet{lee00} and \citet{rey00} to search for variations in the
line flux with the continuum. Both studies found that much of the line
flux was constant and unresponsive to large continuum changes down to
timescales as low as perhaps 0.5~ks. \citet{ve01} fit these data on the orbital
timescale of \textit{RXTE} and found variations in the line flux, but
these were also uncorrelated with the continuum. \citet{fab02} found a
similar result using this \textit{XMM-Newton} observation. These
authors subtracted a 10~ks low flux state spectrum from a 10~ks high flux state
one and found that the difference was well fit by a power-law. The
constancy of the \fe\ line flux, which has been seen in other
Seyfert~1s (e.g., NGC~5548; \citealt{chi00}), is hard to understand if
the entire line arises from close to the primary X-ray source.

Analysis of time-resolved \textit{XMM-Newton} spectra using the double
reflection model found that the inner component became more ionized
when \mcg\ was brighter (the left panel of
Fig~\ref{fig:xi-gam-cr}). Recall from Fig.~\ref{fig:dblfs} that the
spectrum between 2.5 and 12~\kev\ is dominated by emission from the
inner zone. Also, the shape of an ionized reflection spectrum changes
with $\xi$ \citep{ros99}. Thus, the change in ionization parameter
found in the inner reflector is due to variations in the shape of the
spectrum, and not just the \fe\ line. This is not necessarily an
obvious result. \mcg\ softens as it becomes brighter (see the right
panel of Fig.~\ref{fig:xi-gam-cr}), and a steeper illuminating
power-law will decrease the ionization state of the reflector at a
constant irradiating flux. Therefore, in order for $\xi$ to increase
the incident flux must grow at a rate fast enough to offset the
changes in the spectral slope.

As the ionized reflector is nearly coincident with the X-ray source in
 this model, reverberation would be large on short timescales,
 adjusting the shape of the line. Therefore, assuming a constant $R$,
 the changes in the \fe\ line profile observed by \textit{ASCA} can be
 explained by changes in the ionization state of the inner reflector
 \citep[see also][]{rey00}. When the source is bright, the
 reprocessing zone becomes highly ionized and the amplitude of the red
 wing diminishes. On the other hand, when the source is in a low
 state, the ionization parameter drops and there is more emission from
 the red wing explaining the results of \citet{iwa96}. However, can
 the blue core from the distant, neutral reflector respond on the
 correct timescale to explain its variations?  The best fit inner
 radius for the outer reflector is 71~$r_g$
 (Table~\ref{table:dbldata}). This corresponds to a light crossing
 time of 3.5$M_7$~ks, where $M_7=M/10^7$~M$_{\odot}$. \citet{iwa99}
 saw no blue core in their 1997 \textit{ASCA} flare spectrum which
 lasted about an hour -- the same timescale estimated
 above. Therefore, increasing the black hole mass by only a small
 amount will comfortably explain this observation. Yet \citet{fab02}
 found that the difference between a high-state and low-state spectrum
 of \mcg\ is consistent with a power-law. The conclusion from that
 test is that the \fe\ line must not have changed strength or shape
 between the two states.

\begin{figure}
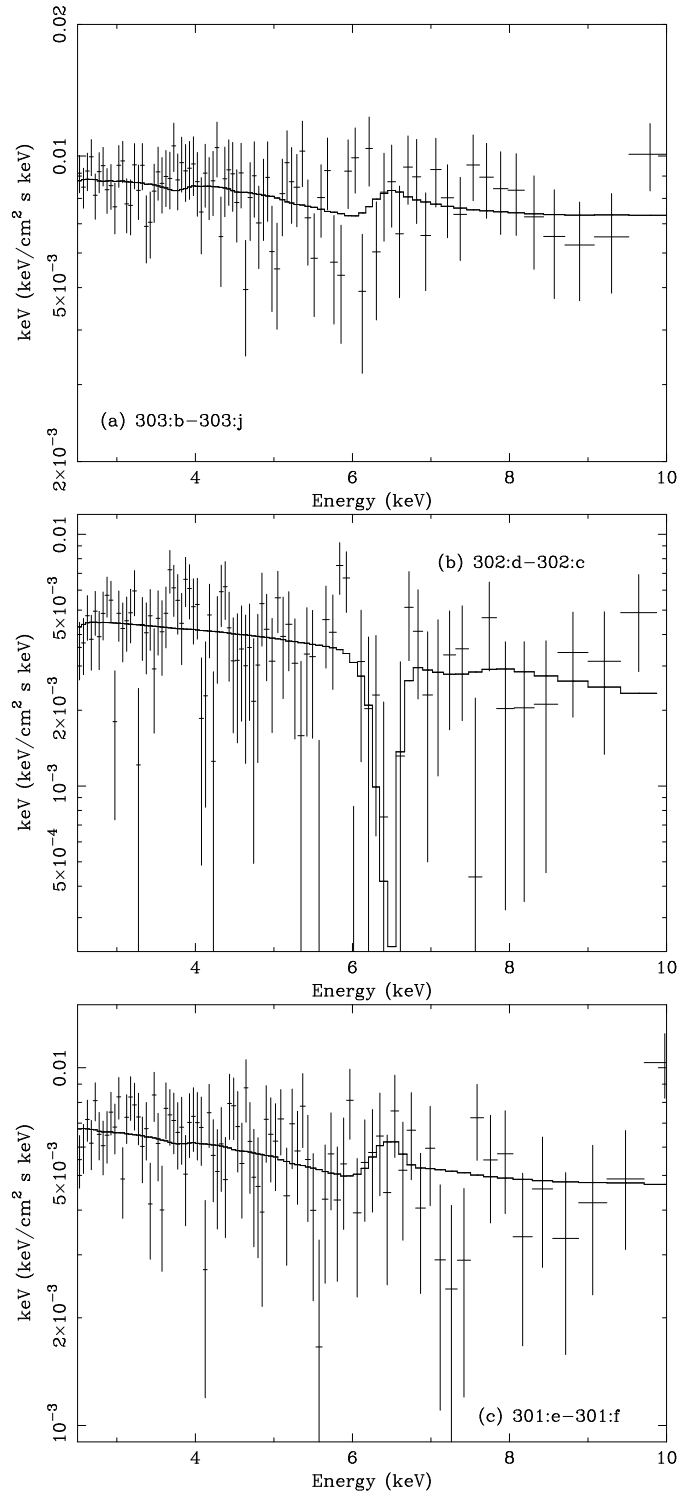

\includegraphics[angle=-90,width=0.5\textwidth]{303b-303j.eps}
\includegraphics[angle=-90,width=0.5\textwidth]{302d-302c.eps}
\includegraphics[angle=-90,width=0.5\textwidth]{301e-301f.eps}
\caption{Unfolded spectra obtained from subtracting the best-fitting double
reflector model from a low flux state interval from the model of a high
state segment. (a) The result when segment Rev. 303:j and 303:b (see
Fig.~\ref{fig:lc}) were used as the low and high flux states,
respectively. These segments are separated by 70~ks. (b) The
result when segment Rev. 302:c and 302:b were used as the low and high
states. These intervals are adjacent to each other. (c) The
result when segment Rev. 301:f and 301:e were used as the low and high
states. These intervals are also adjacent to each other. See the text
for the discussion of the interpretation.}
\label{fig:differences}
\end{figure}

To investigate the line variability predicted by the ionized disc model more
 closely, we show in Figure~\ref{fig:differences} three
 examples of a difference spectrum obtained by subtracting the
 best-fitting double reflector model for a low flux state segment from the
 model of a high-state segment.
Dips in these spectra correspond to regions where the feature was
stronger in the low flux state than in the high flux state, and bumps
correspond to areas where the high flux model was stronger than the
low flux model. The difference spectra obtained from the data were
unfolded and plotted in Fig.~\ref{fig:differences} to
illustrate their statistical quality.

The bottom panel shows the difference between the models for
Rev. 301:e and 301:f (see Fig.~\ref{fig:lc}). At the end of 301:e
there is sudden drop from a 2--10~\kev\ flux of $5.21 \times
10^{-11}$~erg~cm$^{-2}$~s$^{-1}$ to $3.07 \times
10^{-11}$~erg~cm$^{-2}$~s$^{-1}$ in $\sim 1000$~s. The ionization
parameter of the inner reflector fell from $\log \xi = 3.96$ to $3.61$
over that interval, as did the photon index: $\Gamma=1.98$ to
$1.87$. The overall negative slope of the difference spectrum is due
to that change in photon index. At the energy of the red wing of the
\fe\ line there is a slight deficit with respect to the
continuum. This implies that the red wing was stronger in the low flux
state spectrum (as a result of the lower $\xi$) than in the high flux
state spectrum. This feature is weak and very broad and is difficult
to detect in the data. The difference spectrum also shows a hump at
the position of the line core. This means that the 6.4~\kev\ core was
weaker in the low flux state than in the high flux state. If the
distant reflector had equilibrated at the high flux state in segment
301:e, it would have not yet responded to the rapid decrease in flux
unless the black hole mass was $few \times 10^6$~M$_{\odot}$;
alternatively, it may be reacting to changes before
the flare in 301:e, as it lasted only 10~ks (see Fig.~\ref{fig:lc}).

The middle panel of Fig.~\ref{fig:differences} plots the difference
between Rev. 302:d and 302:c, which illustrates the case of the source
moving from a ``steady'' low-state (Rev. 302:a,b,c all have roughly
the same flux) to a bright flare that occurs at the end of the
Rev. 302:d.  Fig.~\ref{fig:lc} shows that the pn count rate increased
by nearly a factor of 2 over the 10~ks interval. The difference
spectrum of these two segments shows a large (factor of 10) drop at
$\sim 6.5$~\kev\ implying that the core of the \fe\ line was stronger
in 302:c than in 302:d, opposite to what is expected if the line core
was responding to the continuum on timescales of $\sim 10$~ks.

The top panel of Fig.~\ref{fig:differences} shows the difference
spectrum between two segments which are separated by 70~ks. The
ionization parameter for the data of Rev. 303:b is $\log \xi = 3.95$
while it is $\log \xi = 3.57$ for segment 303:j. Thus, the red wing of
the \fe\ line, which originates from the ionized reflector should be
stronger in the low flux state spectrum. This does seem to be the
case, as the difference spectrum shows the profile of the red wing as
a deficit between 3 and 6.4~\kev\ which means that this part of the
spectrum is stronger in the low flux state spectrum than in the high
flux state. The core of the line is marginally weaker in the low state
than in the high flux state, consistent with the lower flux
level. Yet, both of these features are $<$10\% from the continuum, and
again, as seen from the data points, difficult to detect
unambiguously. The 2-sigma upper-limit on the EW of any Gaussian
feature at 6.5~keV in the data difference spectrum of 303:b and 303:j
is 120~eV (it is 230~eV for 301:e and 301:f).

The conclusion from these tests is that this model does predict
changes in the line profile on short timescales, in particular around
the red wing. However, these changes will be slight due to the breadth
of the feature and that is why they have not been detected
before. Unfortunately, higher signal-to-noise data at small timescales
will be needed to check these predictions. Finally, the model also
finds that small changes in the 6.4~\kev\ line core are consistent with
the data, but we are unable to determine a timescale for the
changes. Such variations have not been seen before
\citep[e.g.,][]{rey00}, and will need to be found for this multiple
reflection model to be viable, although the signature may be
complex. For example, a warped accretion disc is subject to
Lense-Thirring precession, and this would cause variability in the
outer reflector. Although a numerical simulation is needed
to fully model the dynamics of such systems \citep{an99}, the
precession period can be estimated from \citep{bp75}
\begin{equation}
t_{\mathrm{prec}} \approx 5 \times 10^7 \left({a \over 0.998}
\right)^{-1} \left( {M \over 10^7~\mathrm{M_{\odot}}} \right) \left( {
    r \over 70~r_g} \right)^3\ \ \mathrm{s},
\label{eq:tprec}
\end{equation}
where $a$ is the black hole spin parameter. Therefore, precession can
only cause long timescale variability in this model.

\subsubsection{The inner annulus}
\label{subsub:annul} 
Thus far we have ignored the problem that this model predicts that
the primary X-ray emission
originates from within a tiny annulus at 5~$r_g$. The fit converged to
this value by fitting a notch in the spectrum at $\sim$3.6~\kev; in
order for this to remain unblurred, the outer radius was forced to be
close to the inner radius, which is constrained by the width of the
red wing. Replacing the pn data with the MOS data does not affect this
result. It is possible that this feature is an emission line from
hydrogenic Argon from the outer reflector which is not included in the
reflection models. Further work needs to be done to eliminate this
possibility.

The requirement for the emission to be constrained to an annulus is
very extreme as viscous dissipation throughout the accretion disc
should go as $r^{-3}$ \citep{ss73}. If most of the accretion energy is
transported to a magnetic corona above the disc then it could only be
efficient in this small annulus which seems highly unlikely given that
magnetic fields will be present throughout the flow. In the 1997 flare
spectrum reported by \citet{iwa99}, they also found that the line
emission seemed to arise from an annulus around 5~$r_g$. One
explanation considered by these authors was that this emission was
caused by a flare rotating along the very inner edge of the disc. But,
as these authors pointed out, this implied a black hole mass greater
than 10$^8$~M$_{\odot}$ as well as theoretical problems about the
height of the flare. In the model considered here, emission from an
annulus at 5~$r_g$ is required at nearly every point in the
observation, and cannot be due to a transient event such as a flare.

One possible explanation for this energy generation is a connection
between a spinning black hole and the accretion disc
\citep[e.g.][]{bz77,li00,li02a,li02}. This was considered by \citet{wil01}
to explain the steep emissivity profile found in an earlier
\textit{XMM-Newton} observation of \mcg. Since emission is required
from within 6~$r_g$ it is likely that the black hole is spinning, and
is therefore a source of energy that can be transformed into
radiation. Returning radiation is another method to increase the
emissivity of the inner disc \citep{mmk02,fv02}.

\section{Conclusions}
\label{sect:concl}
Making progress in the study of accretion discs and AGN can only be
made by comparing models with data. Here, we have used the
sophisticated ionized disc models of \citet{ros93} to fit the latest
\textit{XMM-Newton} and \textit{BeppoSAX} spectrum of \mcg. An ionized
disc has often been considered as a possible explanation for the
strange variability behavior of the broad \fe\ line.

We found that an ionized disc model can fit the data and account for
the spectral variability if there are two distinct reflection regions
on the disc. This model requires a very specific geometry in order to
be viable, but can be tested by searching for soft X-ray emission
lines that are broadened to a few 10s of eV, but are not asymmetric,
and, in the future, by the continuum variability properties above
30~\kev. An apparent problem of this scenario is that it constrains
$\sim$80\% of the total X-ray flux of \mcg\ to be produced within a
very narrow annulus only 5~$r_g$ from the black hole. This is a
stringent requirement that is difficult to explain with standard AGN
theory. Nevertheless, the advantage of the double reflection model is
that it can naturally explain the strange \fe\ line variability that
has been previously observed. Time-resolved spectral fitting found
that the inner reflector becomes more ionized as the source brightens,
weakening the red wing of the iron line. Changes in the 6.4~\kev\ line
core are also found in the time-resolved fitting, but they do not
necessarily correlate with the continuum. The model can also account
for the $R > 2$ constraint implied by the \sax\ data by combining the
reflection humps of multiple reprocessors. 

One similarity that the model presented here has with previous work
\citep{wil01,fab02} is the need for extra production of
radiation within 6~$r_g$. Therefore, irrespective of the number of
reflectors or the ionization state of the disk, it seems likely there
is some new physical process near the black hole of \mcg\ that needs
to be untangled.
 
\section*{Acknowledgments}
Based on observations obtained with \textit{XMM-Newton}, an ESA
science mission with instruments and contributions directly funded by
ESA Member States and the USA (NASA).  DRB acknowledges financial
support from the Commonwealth Scholarship and Fellowship Plan and the
Natural Sciences and Engineering Research Council of Canada. ACF
thanks the Royal Society for support. We thank the anonymous referee
for useful comments.


\bsp 

\label{lastpage}

\end{document}